\documentclass[preprint,10pt]{elsarticle}

\usepackage{latexsym, amsmath, amsfonts, amssymb, amsthm, graphicx, epsfig, epstopdf}
\usepackage{array, longtable, calc, multirow, hhline} 
\usepackage[lined,boxed,linesnumbered]{algorithm2e}
\usepackage{mathrsfs}

\def\b{\mathbf b}
\def\d{\mathbf d}
\def\e{\mathbf e}
\def\f{\mathbf f}
\def\r{\mathbf r}
\def\1{\mathbf 1}
\def\0{\mathbf 0}
\def\A{\mathbf A}
\def\I{\mathbf I}
\def\L{\mathcal L}
\def\U{\mathcal U}
\def\V{\mathit V}
\def\X{\mathfrak X}

\newtheorem{Example}{Example}

\begin{document}
	
\begin{frontmatter}

\title{Heuristic construction of exact experimental designs under multiple resource constraints}
\author{Radoslav Harman\corref{cor1}}
\ead{harman@fmph.uniba.sk}
\author{Alena Bachrat\'{a}\corref{cor2}}
\author{Lenka Filov\'{a}\corref{cor3}}
\cortext[cor1]{Corresponding author. Tel.: +421 2 602 95 717}

\address{Department of Applied Mathematics and Statistics, Faculty of Mathematics, Physics and Informatics, Comenius University, Mlynsk\'{a} dolina, 84248 Bratislava, Slovakia}

\begin{abstract}
   
The aim of this paper is twofold. First, we introduce ``resource constraints'' as a general concept that covers many practical restrictions on experimental design. Second, for computing efficient exact designs of experiments under any combination of resource constraints, we propose a tabu search heuristic that uses some ideas of the Detmax procedure. To illustrate the scope and performance of our heuristic, we computed $D$-efficient designs for 1) a block model with limits on the numbers of blocks and on the availability of experimental material; 2) a quadratic regression model with simultaneous marginal and cost constraints; 3) a non-linear regression model with simultaneous direct and cost constraints. As we show, the proposed heuristic generates comparable or better results than algorithms specialized for computing optimal designs under less general constraints.
\end{abstract}
   
\begin{keyword}
	Design of experiments \sep $D$-optimality \sep Heuristic optimization \sep Tabu search \sep Detmax procedure
	\MSC[2010] 62K05
\end{keyword}

\end{frontmatter}

\section{Introduction}\label{sec:Introduction}

Optimal design of experiments is an approach to constructing experimental designs using a statistically motivated utility function called an optimality criterion, see, e.g., \cite{Pazman}, \cite{FedorovHackl}, \cite{Pukelsheim} and \cite{Atkinson}. Construction of an optimal experimental design is generally a challenging problem of theoretical mathematics and numerical optimization. In this paper, we propose a unifying view on various experimental design restrictions encountered in practice, which we formalize by the notion of ``resource constraints''. We show that efficient designs under any system of resource constraints can be constructed by a single heuristic method.

Suppose that we intend to perform an experiment consisting of a set of trials (runs, measurements). For each trial, we must select a design point from a finite design space $\mathfrak{X}$ representing permissible experimental conditions. We assume that in general it is possible to select the same design point for more than one trial.

 For each $x \in \mathfrak{X}$, let $\iota(x)$ be the index of $x$, i.e., $\iota$ is a one-to-one mapping from $\mathfrak{X}$ to $\{1{:}n\}:=\{1,...,n\}$, where $n$ is the size of $\mathfrak{X}$. In this setting, an ``exact'' experimental design can be represented by a vector $\xi \in \{0,1,2,...\}^n$ with components $\xi_1,....,\xi_n$ determining the numbers of independently replicated trials in the design points $\iota^{-1}(1),...,\iota^{-1}(n)$, respectively. A vector $\xi \in [0,\infty)^n$ with general non-negative components will be called an ``approximate'' experimental design, and viewed as a relaxation of an exact design\footnote{Note that we do not use the definition of experimental designs as a probability (that is, normalized) measures on $\X$. The reason is that in the problems with multiple resource constraints the number of trials (i.e., the normalization constant) is not known in advance.}.

The following toy example motivated by Question 2.2 in \cite{Bailey-book} will be used to illustrate the basic definitions. 

\begin{Example}[]\label{Example}
An engineer wants to protect metal plates against corrosion. There is a new paint for the plates. The engineer decides to estimate the protective effect of one and two coats of the new paint. He will paint some metal plates once, some twice, and then immerse them all in a tank of water. Later he will remove all the plates, and measure the degree of corrosion of each. In this situation, the design space is $\mathfrak{X}=\{\textrm{one coat}, \textrm{ two coats}\}$ with a natural indexing $\iota: \mathfrak{X} \to \{1,2\}$. An exact experimental design is any vector $\xi=(\xi_1,\xi_2)^T$, where $\xi_1$ means the number of metal plates painted with one coat and $\xi_2$ means the number of metal plates painted with two coats. An approximate experimental design is any two-dimensional vector with non-negative components.
\end{Example}

Let $\phi: [0, \infty)^n \to [0, \infty)$ be an optimality criterion that measures the quality of (exact or approximate) designs for statistical inference. Often, the goal of the experimenter is to estimate unknown parameters of an underlying statistical model, and the value $\phi(\xi)$ is a measure of the information about the parameters of interest obtained from the experiment $\xi$, see, e.g., Chapter 5 in \cite{Pukelsheim}. In view of this interpretation, it is natural to adopt the following assumption of monotonicity:
\begin{itemize}
  \item[(M)] Augmentation (extension) of an experiment by additional trials cannot decrease its quality for statistical inference, i.e., if designs $\xi$ and $\zeta$ satisfy $\xi \leq \zeta$ componentwise, then $\phi(\xi) \leq \phi(\zeta)$.
\end{itemize}
The most classical example of $\phi$ is the criterion of $D$-optimality for linear regression models with independent homoscedastic errors, as we will briefly describe.

Consider an experiment with an $n$-point design space $\mathfrak{X}$. Assume that for each trial in the design point $x \in \mathfrak{X}$, the real-valued random observation $Y$ satisfies $E(Y)=\f_{\iota(x)}^T\beta$ and $\mathrm{Var}(Y) = \sigma^2 < \infty$, where $\f_{\iota(x)} \in \mathbb{R}^m$ is a ``regressor'' vector, $\iota(x) \in \{1{:}n\}$ is the index of $x$, $\beta \in \mathbb{R}^m$ is a vector of unknown parameters of interest, and $\sigma^2$ is a constant variance. For different trials, the observations are assumed to be independent. Then, the criterion of $D$-optimality $\phi_D: [0,\infty)^n \to [0,\infty)$ is defined by 
\begin{equation*}
  \phi_D(\xi)=[\det(\sum_{i=1}^n \xi_i \f_i\f^T_i)]^{1/m}.
\end{equation*}
It is possible to show that $\phi_D$ is continuous, concave and monotonic on $[0,\infty)^n$ in the sense of Assumption (M); see, e.g., Chapter 5 and Section 6.2 in \cite{Pukelsheim}. Additionally, $\phi_D$ is homogeneous, that is, $\phi_D(c\xi)=c\phi_D(\xi)$ for any design $\xi$ and any $c \geq 0$. Hence, the quality of two designs can be compared by their relative $D$-efficiency defined by $\mathrm{eff}_D(\xi|\zeta)= \phi_D(\xi)/\phi_D(\zeta)$ for all $\xi$ and $\zeta$ such that $\phi_D(\zeta)>0$. A design $\xi^*$ is called $D$-optimal, if it maximizes the value of $\phi_D$ in a given set $\Xi$ of competing designs. From the statistical point of view, the $D$-optimal design minimizes the generalized variance of the best linear unbiased estimator of $\beta$ or, in the case of normal observations, a confidence ellipsoid for $\beta$. For details, see \cite{Pazman}, \cite{FedorovHackl}, \cite{Pukelsheim}, and \cite{Atkinson}. 

\setcounter{Example}{0}
\begin{Example}[continued]
 A possible model for observations in $x \in \mathfrak{X}$ (that is, for measurements of the degree of corrosion) is $E(Y)=\f_{\iota(x)}^T\beta$, $\mathrm{Var}(Y) = \sigma^2 < \infty$. Here, $\f_1=(1,0)^T$, $\f_2=(0,1)^T$, and $\beta=(\beta_1,\beta_2)^T$, i.e., $\beta_1$ is the mean value of the degree of corrosion with one coat and $\beta_2$ is the mean value of the degree of corrosion with two coats. For a design $\xi=(\xi_1,\xi_2)^T$ the value of the $D$-criterion is $  \phi_D(\xi)=\sqrt{\xi_1\xi_2}$. 
\end{Example} 

Usually, the set of designs is only restricted by the number of trials, i.e., 
\begin{equation}\label{standard}
  \xi_1+...+\xi_n \leq N
\end{equation}
for each feasible exact design $\xi$, where $N$ is a maximum ``size'' of the experiment.\footnote{In fact, a more common requirement is that the number of trials is \emph{exactly} equal to $N$, but assumption (M) implies that this requirement is equivalent to \eqref{standard}.} This corresponds to the situation where each trial amounts to the same cost and the experimental budget allows performing at most $N$ trials. Alternatively, condition \eqref{standard} can represent the requirement that the trials must be performed sequentially, each trial lasts the same amount of time, and the deadline permits performing at most $N$ trials. However, there are practical situations where any feasible design must satisfy one or more constraints different from \eqref{standard}.

\setcounter{Example}{0}
\begin{Example}[continued]
 In our example, a natural restriction is that the number of available metal plates is $N$, that is, we can select only designs $\xi$ that satisfy $\xi_1+\xi_2 \leq N$. However, even in this extremely simple case, there might be some additional constraints. For instance, it is possible that the total available amount of the paint is limited by $b$ units and one coat of the paint consumes $a_c$ units. In this case, we must add a second restriction $a_c\xi_1+2a_c\xi_2 \leq b$ on feasible experimental designs $\xi$. 
\end{Example} 

The study of experimental designs under multiple constraints\footnote{Note that in this paper we consider the constraints on the experimental design itself, which are different from the constraints on the design \emph{space} that are also relevant to some applications; see, for instance, Section 12.7 in \cite{Atkinson}.} is an important part of optimal design theory, see, for instance, the review paper \cite{CookFedorov} or Chapter 4 in \cite{FedorovHackl}. Generally, the constrained design problems are difficult, especially for a large design space. For instance, even in the case of a linear system of constraints, mere finding a single feasible exact design (or proving that there is no such design) may be a highly non-trivial task. In this paper, we introduce a class of constraints that do not comprise all linear constraints on designs, yet they are broad enough to encompass many practical experimental design restrictions and, at the same time, lead to a relatively simple set of feasible exact experimental designs.

We propose to consider general ``resource'' constraints of the form
\begin{equation}\label{resconst}
  \sum_{i=1}^n a_{ri} \xi_i \leq b_r \text{ for all } r \in \{1{:}k\},
\end{equation}  
where $a_{ri}$ represents the consumption of the $r$-th resource by a single trial in the $i$-th design point, and $b_r$ represents a limit on the $r$-th resource. 

 The interpretation of \eqref{resconst} leads to the following assumptions:
\begin{itemize}
  \item[(C1)] Resource limits are positive and finite, i.e., $b_1,...,b_k \in (0, \infty)$.
  \item[(C2)] Augmenting designs can not decrease the consumption of any resource, i.e., $a_{ri} \geq 0$ for all $r \in \{1{:}k\}$ and $i \in \{1{:}n\}$.
  \item[(C3)] No trial is completely free and its replication must eventually result in exceeding some resource limit, i.e., for all $i \in \{1{:}n\}$ there is some $r \in \{1{:}k\}$ such that $a_{ri} > 0$.
\end{itemize}
Clearly, constraint \eqref{standard} is a special case of \eqref{resconst} with $k=1$, $a_{1i}=1$ for all $i \in \{1{:}n\}$, and $b_1=N$. Sometimes, however, the costs depend on design points and the budget of the experiment is limited by $B$ financial units (e.g., Section 6 in \cite{Elfving},  \cite{ParkMontgomery}, or \cite{Wright}). This restriction can be formalized by one resource constraint ($k=1$) such that $b_1=B$, and $a_{11},...,a_{1n}$ represent possibly unequal costs of trials in individual design points. 

A natural type of restrictions are the so-called direct constraints (e.g., \cite{Fedorov82}, \cite{Welch84}), which correspond to performing at most $l_1,...,l_n$ trials in design points $\iota^{-1}(1),...,\iota^{-1}(n)$, respectively. Often, it is possible to perform at most one observation in any design point, i.e., $l_i=1$ for all $i \in \{1:n\}$, as in \cite{Wright}. This can be converted to the resource constraints \eqref{resconst} by setting $k=n$, $a_{ri}=\delta_{ri}$ (the Kronecker delta) and $b_r=l_r$ for all $r \in \{1{:}k\}$. 

 Another class of constraints corresponds to the so-called marginal restrictions (e.g., \cite{CookThibodeau}, \cite{MartinMartin}) or, more generally, strata restrictions (\cite{Harman14}). In this case, design space $\mathfrak{X}$ is partitioned into non-overlapping sets $\mathfrak{X}_1,...,\mathfrak{X}_k$ and any experimental design $\xi$ must satisfy $\sum_{x \in \mathfrak{X}_r} \xi_{\iota(x)} \leq s_r$ for all $r \in \{1{:}k\}$, where $s_1,...,s_k$ are given positive numbers. Here, we obtain \eqref{resconst} by setting $a_{r\iota(x)}=I[x \in \mathfrak{X}_r]$ (the indicator function) and $b_r=s_r$ for all $r \in \{1{:}k\}$.
 
The resource constraints can also accommodate general limits on the availability of experimental material, such as treatment samples in block designs (cf. \cite{EYSM}): Consider some selection $\mathfrak{X}_1,...,\mathfrak{X}_k$ of subsets of $\mathfrak{X}$ and assume that for a trial in any design point $x \in \mathfrak{X}_k \subset \mathfrak{X}$, one piece of experimental material from a pool of $s_k$ available pieces is consumed. This leads to the resource constraints with $a_{r\iota(x)}=I[x \in \mathfrak{X}_r]$ and $b_r=s_r$ for all $r \in \{1{:}k\}$, similarly to the stratified designs.

Yet another type of constraint is the requirement that the trials should not be in ``close'' design points. For instance, if $\mathfrak{X}=\{1:T\}$ represents a sequence of time moments of trials, then the experimenter may be faced with a technical requirement that consecutive trials must be at least $\Delta$ ($\leq T$) time moments apart. This is also possible to express using $k=T-\Delta+1$ resource constraints by setting $a_{ri}=I[i \in \{r,r+1,...,r+\Delta-1\}]$ for all $r \in \{1{:}k\}$, $i \in \{1:n\}$, and $b_r=1$ for all $r \in \{1{:}k\}$. See Section 5 in \cite{SagnolHarman} for an example. 

However, we remark that there do exist some reasonable experimental design restrictions that can not be represented by \eqref{resconst}, for instance non-standard equality constraints on some of the values $\xi_1,...,\xi_N$, or limits on transitional costs (cf. \cite{TackVandebroek}).
\bigskip

Let $\xi^{(0)}$ be either an exact design representing trials that have already been performed, or a required initial part of the experiment. We will assume that $\xi^{(0)}$ satisfies \eqref{resconst}. In the situation without existing/required trials, the design $\xi^{(0)}$ is simply the zero vector $\mathbf{0}_n \in \mathbb{R}^n$.

In matrix form, the system \eqref{resconst} can be written as $\A\xi \leq \b$ componentwise, where $\A$ is the $k \times n$ matrix of coefficients $a_{ri}$, $r \in \{1{:}k\}$, $i \in \{1{:}n\}$, and $\b$ is the $k$-dimensional vector of $b_1,...,b_k$. The assumptions (C1)-(C3) guarantee that the set
\begin{equation*}
 \Xi^{\mathrm{ex}}=\{\xi \in \{0,1,2,...\}^n: \xi^{(0)} \leq \xi, \A\xi \leq \b\}
\end{equation*}
of all feasible exact experimental designs is non-empty and finite. 
Although general enough to represent many practical restrictions on experimental design, set $\Xi^{\mathrm{ex}}$ is still relatively simple to be explored by heuristic optimization methods based on transitions between ``neighbouring'' feasible solutions.

The assumptions also imply that the set of all feasible approximate designs
\begin{equation*}
  \Xi^{\mathrm{ap}}=\{\xi \in [0,\infty)^n: \xi^{(0)} \leq \xi, \: \A\xi \leq \b\}
\end{equation*}
is a non-empty, compact and convex polyhedron. 

The purpose of this paper is to develop a method for solving the general resource-constrained exact optimal design problem
\begin{equation}\label{GeneralProblem}
  \xi^* \in \mathrm{arg}\max\{\phi(\xi): \xi \in \Xi^{\mathrm{ex}}\}.
\end{equation}

\setcounter{Example}{0}
\begin{Example}[continued]
 If we assume $\xi^{(0)}=(0,0)^T$, set $\Xi^{\mathrm{ap}}$ is the polygon $\{(\xi_1,\xi_2)^T: \xi_1 \geq 0, \xi_2 \geq 0, \xi_1+\xi_2 \leq 1, a\xi_1+2a\xi_2 \leq b\}$ and set $\Xi^{\mathrm{ex}}$ is the intersection of $\Xi^{\mathrm{ap}}$ with the integer lattice. If we have, for instance, $N=20$, $a=1$, and $b=23$, then, the globally $D$-optimal exact design found by the complete enumeration is $\xi^*=(11,6)^T$, that is, $11$ plates should be painted with one coat and $6$ plates should be painted with two coats. Although the problem is very small, the $D$-criterion has as many as $5$ strict local optima\footnote{These local optima are $(9,7)^T$, $(11,6)^T$, $(13,5)^T$, $(15,4)^T$, $(17,3)^T$} on $\Xi^{\mathrm{ex}}$ if, for each exact design $\xi$, we allow transitions to all natural neighbours of $\xi$\footnote{These neighbours have the form $(\xi_1+\delta_1,\xi_2+\delta_2)^T$, where $\delta_1,\delta_2 \in \{-1,0,+1\}$, $(\delta_1,\delta_2)^T \neq \0_2$.} that belong to $\Xi^{\mathrm{ex}}$.
\end{Example} 

Besides optimum design, optimization \eqref{GeneralProblem} covers many other difficult discrete optimization problems, for instance knapsack problems (e.g., \cite{Hochbaum}, \cite{Kellerer}), optimal redundancy allocation in reliability theory (e.g., \cite{Chern}, \cite{Kuo}), and constructing $t$-optimal graphs (see Section \ref{Dblock} for more details).

For small to medium size problems of type \eqref{GeneralProblem}, it is possible to use an ``intelligent'' enumeration method, such as branch-and-bound or branch-and-cut, that guarantees a globally optimal solution (see \cite{Welch82}, \cite{SagnolHarman}, cf. \cite{HarmanFilova}). Nevertheless, there is no practical hope of creating an algorithm that rapidly produces provably optimal solutions of large instances of \eqref{GeneralProblem}. Often, the only possibility is to use a heuristic that usually leads to an efficient feasible experimental design. 

A natural approach to solving \eqref{GeneralProblem} is to use a heuristic based on ``excursions'' within the set of designs, as in some early algorithms for computing $D$-optimal experimental designs under the standard constraint \eqref{standard}, see, e.g. \cite{Dykstra}, \cite{Wynn} and \cite{Mitchell}. From these algorithms, the most relevant to our problem is the Detmax procedure proposed by Mitchell (\cite{Mitchell}), which is related to the tabu search methods (e.g., \cite{GloverLaguna}, see also \cite{JungYum}).

Today, the most popular methods for solving the standard optimal design problems are local-search exchange heuristics (e.g., Chapter 12 in \cite{Atkinson}, and \cite{Jones}). However, the exchange heuristics cannot be directly used to solve instances of the general problem \eqref{GeneralProblem}, since the number of trials of the optimal resource constrained design is not known in advance and, in addition, an exchange of two design points may render a feasible design non-feasible. Moreover, our experience shows that problems with resource constraints tend to have a large number of local optima (cf. Example \ref{Example}), i.e., an efficient modification of exchange heuristics requires means for overcoming their local-search nature. 

As far as more general constraints are concerned, an idea similar to Detmax has already been used in reliability theory to efficiently solve redundancy optimization problems (see \cite{KimYum}). In the area of optimal design of experiments, the paper \cite{Welch84} studied a modification of the Detmax procedure capable of computing optimal designs under direct constraints. In \cite{Wright}, another related procedure has been developed, based on a sequential removal of a single design point and a subsequent augmentation of the resulting design by a greedy method. However, this method is restricted to solving optimum design problems with particular kind of direct constraints combined with a single cost constraint. Finally, the paper \cite{EYSM} describes a randomized algorithm similar to simulated annealing that can be used to solve the general problem \eqref{GeneralProblem}. This algorithm had been a starting point of the development of the heuristic proposed in this paper. 


\section{Heuristic}\label{sec:Algorithm}

\subsection{General description of the heuristic}

We will say that a design $\zeta \in \Xi^{\mathrm{ex}}$ is created from a design $\xi \in \Xi^{\mathrm{ex}}$ by a forward step (or a backward step) if $\zeta = \xi +\e_i$ (or $\zeta = \xi -\e_i$) for some standard unit vector $\e_i \in \mathbb{R}^n$. A design $\xi \in \Xi^{\mathrm{ex}}$ will be called ``maximal'' if it can not be augmented without violation of some of the resource constraints, that is, if all designs created from $\xi$ by a forward step are non-feasible. 

For a design $\xi \in \Xi^{\mathrm{ex}}$, an upper neighbour is any feasible design that can be obtained from $\xi$ by a forward step, i.e., the set of all upper neighbours of $\xi$ is
\begin{equation*}
 \U(\xi)=\{\xi+\e_1,...,\xi+\e_n\} \cap \Xi^{\mathrm{ex}}.
\end{equation*}
Similarly, a lower neighbour of a feasible design $\xi$ is any design that can be obtained from $\xi$ by a backward step, that is, the set of all lower neighbours of $\xi$ is \begin{equation*}
  \L(\xi)=\{\xi-\e_1,...,\xi-\e_n\} \cap \Xi^{\mathrm{ex}}.
\end{equation*}
Note that $\L(\xi^{(0)})=\emptyset$, and $\U(\xi)=\emptyset$ if and only if $\xi$ is maximal. We will also assume that $\xi^{(0)}$ is not maximal, which means that $\L(\xi) \cup \U(\xi) \neq \emptyset$ for any feasible design $\xi$.

Clearly, properties (C1)-(C3) imply that any feasible exact design is reachable from any other feasible exact design by a sequence of forward and backward steps within $\Xi^{\mathrm{ex}}$. Moreover, an optimal solution of \eqref{GeneralProblem} can be found among maximal designs, in view of assumption (M). 

The proposed algorithm starts in a design $\xi^{(1)} \in \Xi^{\mathrm{ex}}$ (cf. Subsection \ref{Subsec:ini}) and builds an excursion in the set of feasible exact designs, guided by a ``tabu'' list $\V$ of characteristic attributes (e.g., numeric identifiers) of the designs that have already been visited.  

Let $\mathit{attr}(\xi)$ be a characteristic attribute of $\xi\in \Xi^{\mathrm{ex}}$ and let $\mathit{val}(\xi)$ be a local heuristic evaluation of $\xi$, i.e., a real number that roughly estimates how promising $\xi$ is as a part of an excursion leading to an efficient design (see Subsections \ref{Subsec:attr} and \ref{Subsec:val} for a more detailed specification of $\mathit{attr}$ and $\mathit{val}$).

Let $\xi$ represent the current design in the excursion. The algorithm first attempts a forward step (if $\mathit{attr}(\xi) \notin \V$) or a backward step (if $\mathit{attr}(\xi) \in \V$), moving to a neighbouring feasible exact design $\zeta$. Design $\zeta$ is chosen such that it maximizes  $\mathit{val}$ among all designs satisfying $\mathit{attr}(\zeta) \notin \V$. If the algorithm attempts a forward step but there is no $\zeta \in \U(\xi)$ such that $\mathit{attr}(\zeta) \notin \V$, or if it attempts a backward step but there is no $\zeta \in \L(\xi)$ such that $\mathit{attr}(\zeta) \notin \V$, the algorithm tries to reverse the direction of the search. If all these attempts fail, i.e., if the attributes of all neighbouring designs of $\xi$ are contained in the list $\V$, the algorithm resolves this ``blockage'' by randomly selecting a design from $\L(\xi) \cup \U(\xi)$ for the next step.

Each time a maximal design is encountered, the algorithm checks whether it is better than the best feasible design $\xi^+$ found so far. If the number of backward steps of an excursion exceeds a constant $\mathit{back}_{\max}$, the excursion is declared to be a ``failure'' and the algorithm is restarted from the currently best design. Note that the list $\V$ is not cleared after the restart, that is, the new excursion will follow a different path. The algorithm is terminated once the computation time exceeds a user-supplied time limit $\mathit{t}_{\max}$. The idea of the algorithm is made more precise by its meta-heuristic scheme; see Algorithm \ref{Hugo}.

Thus, the algorithm is similar to the Detmax procedure, because it attempts a forward or a backward step depending on an attribute of the current design. Since most of the designs encountered have not been previously visited, the excursions tend to move towards maximal designs. Note, however, that even under the standard constraint, Algorithm \ref{Hugo} differs from the Detmax procedure in several important aspects. For instance, unlike Algorithm \ref{Hugo}, the Detmax procedure does not avoid the backward steps to the designs with attributes in the tabu list, which often leads to retracing the same excursions. Of course, from the point of view of this paper, the main disadvantage of the Detmax algorithm is that it is only suitable for computing optimal designs under the standard constraint on the size of the experiment.

\IncMargin{1em}
\begin{algorithm}
\SetKwFunction{OptimalDesign}{OptimalDesign}
\SetKwInOut{Input}{Input}\SetKwInOut{Output}{Output}
\Input{The model, matrix $\A$ of consumption coefficients, vector $\b$ of resource limits, design $\xi^{(0)}$ to be augmented, initial feasible design $\xi^{(1)}$, required time of computation $\mathrm{t}_{\max}$, criterion $\phi$, maximum number of backward steps $\mathrm{back}_{\max}$.}
\Output{A design $\xi^+$ as the best found feasible solution of
\eqref{GeneralProblem}.} \BlankLine $\xi^+ \leftarrow \xi \leftarrow
\xi^{(1)}$; $\V \leftarrow \emptyset$; $\mathrm{back}_{\mathrm{no}}
\leftarrow 0$\;
\Repeat{$time>\mathrm{t}_{\max}$}{
\eIf{$\mathit{attr}(\xi) \notin \V$}
   {$\V \leftarrow \V \cup \{\mathit{attr}(\xi)\}$\;
    \eIf{$\{\mathit{attr}(\zeta): \zeta \in \U(\xi)\} \nsubseteq \V$}
       {$\xi \leftarrow \mathrm{argmax}\{\mathit{val}(\zeta): \zeta \in \U(\xi), \mathit{attr}(\zeta) \notin \V\}$\;
       }{
         {\If{$\U(\xi)=\emptyset$ $\mathrm{ and }$ $\phi(\xi^+)<\phi(\xi)$}
            {$\xi^+ \leftarrow \xi$;  $\mathrm{back}_{\mathrm{no}} \leftarrow 0$\;}
          \eIf{$\{\mathit{attr}(\zeta): \zeta \in \L(\xi)\} \nsubseteq \V$}
              {$\xi \leftarrow \mathrm{argmax}\{\mathit{val}(\zeta): \zeta \in \L(\xi), \mathit{attr}(\zeta) \notin \V\}$\;
               $\mathrm{back}_{\mathrm{no}} \leftarrow \mathrm{back}_{\mathrm{no}}+1$\;
          }{{$\xi \leftarrow$ a random design $\zeta$ from $\L(\xi) \cup \U(\xi)$}
              }
          }
       }
   }{
      {
       \uIf{$\{\mathit{attr}(\zeta): \zeta \in \L(\xi)\} \nsubseteq \V$}
           {$\xi \leftarrow \mathrm{argmax}\{\mathit{val}(\zeta): \zeta \in \L(\xi), \mathit{attr}(\zeta) \notin \V\}$\;
            $\mathrm{back}_{\mathrm{no}} \leftarrow \mathrm{back}_{\mathrm{no}}+1$\;}
           \uElseIf{$\{\mathit{attr}(\zeta): \zeta \in \U(\xi)\} \nsubseteq \V$}
              {$\xi \leftarrow \mathrm{argmax}\{\mathit{val}(\zeta): \zeta \in \U(\xi), \mathit{attr}(\zeta) \notin \V\}$\;}
           \Else{$\xi \leftarrow$ a random design $\zeta$ from $\L(\xi) \cup \U(\xi)$}
       }
     }
  \lIf{$\mathrm{back}_{\mathrm{no}} > \mathrm{back}_{\max}$}{$\xi \leftarrow \xi^+$; $\mathrm{back}_{\mathrm{no}} \leftarrow 0$}
} 

\caption{A general scheme of the proposed heuristic for computing efficient experimental designs under constraints \eqref{GeneralProblem}. For linear regression, the model at the input can be represented by an $m \times n$ matrix $\mathbf{F}=(\f_1,...,\f_n)$ of regressors corresponding to individual design points. The model at the input is used inside the functions $\phi$, $\mathit{val}$, and can also be used inside $\mathit{attr}$.  Matrix $\A$ and vector $\b$ define the resource constraints. They are implicitly used for computing the set $\U(\xi)$, and can also be used inside $\mathit{val}$.  In our implementation, the function $\mathit{attr}$ used an additional constant $\mathrm{n}_\mathrm{round}$; see Subsection \ref{Subsec:attr}. Design $\xi^{(0)}$ is implicitly used for computing the set $\L(\xi)$. The function $time$ returns the time elapsed from the start of the computation.}\label{Hugo} \end{algorithm}\DecMargin{1em}

\subsection{Choice of the initial design}\label{Subsec:ini}

An important part of Algorithm \ref{Hugo} is the choice of the initial design $\xi^{(1)}$. Our experience shows that a reasonably efficient design can usually be obtained by choosing $\xi^{(1)}=\xi^{(0)}$, but for more complex problems it is better to use multiple restarts of the heuristic, with initial designs created by a sequence of random forward steps starting from $\xi^{(0)}$.
 
Another possibility is to take an optimal approximate design $\tilde{\xi} \in \mathrm{arg}\max\{\phi(\xi): \xi \in \Xi^{\mathrm{ap}}\}$ and set $\xi^{(1)}=(\lfloor \tilde{\xi}_1 \rfloor, ..., \lfloor \tilde{\xi}_n \rfloor)$, where $\lfloor \cdot \rfloor$ denotes the floor function. The nature of the resource constraints guarantees that the design $\xi^{(1)}$ constructed in this way will be feasible. For computing approximate optimal designs under various types of linear constraints, one can use efficient convex optimization methods, see, for instance, \cite{Vandenberghe} and  \cite{SagnolHarman}.

\subsection{Choice of the characteristic attributes of designs}\label{Subsec:attr}

The characteristic attribute should be chosen such that, loosely speaking, it assigns different values to substantially different designs and the same values to essentially same designs (for instance to algebraically isomorphic designs).

After some experimentation, we have decided to use the attribute $\mathit{attr}(\zeta)$ equal to the value $\phi(\zeta)$ rounded to $\mathit{n}_{\mathrm{round}}$ significant digits. Note that instead of storing complete designs, storing real-valued attributes in the list $\V$ not only makes the time and memory requirements much smaller, but sometimes makes the tabu principle itself more efficient. This is the case in models with many statistically isomorphic designs, because including an attribute based on $\phi(\zeta)$ into $V$ has the effect of ``blocking'' also all designs isomorphic with $\xi$.

\subsection{Choice of the local heuristic evaluation of designs}\label{Subsec:val}

In our implementation of Algorithm \ref{Hugo}, the local heuristic evaluation $\mathit{val}(\zeta)$ of a design $\zeta$ is an estimate of the maximal value of $\phi$ on the set of all designs augmenting $\zeta$, i.e., $\mathit{val}(\zeta)$ is an estimate of
\begin{equation*}
\mathit{val}^*(\zeta)=\max\{\phi(\eta): \eta \in \Xi^{\mathrm{ex}}, \zeta \leq \eta \}.
\end{equation*}
The rationale behind this particular evaluation is that if we were able to use the exact values of $\mathit{val}^*$, the initial greedy phase of the algorithm started from $\xi^{(0)}$ would directly lead to a globally optimal solution.

For $\zeta \in \Xi^{\mathrm{ex}}$, let 
\begin{equation*}
  \r(\zeta)=(b_1-\sum_{i=1}^n a_{1i}\zeta_i,...,b_k-\sum_{i=1}^n a_{ki}\zeta_i)^T
\end{equation*}
be the vector of residual amounts of resources. Note that after a forward or a backward step, it is possible to use the update formula
\begin{equation*}
  \r(\zeta \pm \e_i)=\r(\zeta) \mp (a_{1i},...,a_{ki})^T, \ i \in \{1{:}n\}.
\end{equation*}

For every $i \in \{1{:}n\}$, let
\begin{eqnarray}\label{eqdelta}
  d_i(\zeta)&=&\max\{d \geq 0: \zeta+d \e_i \in \Xi^{\mathrm{ex}}\} \nonumber \\
            &=&\left\lfloor \min\left\{a_{ri}^{-1}\r_r(\zeta): r \in \{1{:}k\}, a_{ri}>0\right\} \right\rfloor. 
\end{eqnarray} 

The vector $\d(\zeta)=(d_1(\zeta),...,d_n(\zeta))^T$ estimates the direction towards ``large'' feasible designs. Furthermore, if $\d(\zeta) \neq \0_n$ define 
\begin{eqnarray}\label{eqgamma}         
  \gamma(\zeta)&=&\max\{\gamma \geq 0: \zeta+\gamma \d(\zeta) \in \Xi^{\mathrm{ap}}\} \nonumber \\
  &=&\min\left\{h_r^{-1}(\zeta)\r_r(\zeta): r \in \{1{:}k\}, h_r(\zeta)>0\right\}, 
\end{eqnarray} 
where $h_r(\zeta)=\sum_{i=1}^n a_{ri}d_i(\zeta)$. If $\d(\zeta) = \0_n$, i.e., if $\zeta$ is a maximal design, define $\gamma(\zeta)=0$. The vector  $\zeta+\gamma(\zeta)\d(\zeta)$ is the ``largest'' feasible approximate design in the direction $\d(\zeta)$. Thus, the value
\begin{equation}\label{eq:val}
  \mathit{val}(\zeta)=\phi(\zeta+\gamma(\zeta) \d(\zeta))
\end{equation}
gives us a rough estimate of $\mathit{val}^*(\zeta)$. For some designs $\zeta \in \Xi^{\mathrm{ex}}$ the vector $\zeta+\gamma(\zeta) \d(\zeta)$ may have non-integer components, that is, the proposed heuristic evaluation is based on the criterial values of general approximate designs. 

We remark that formulas \eqref{eqdelta} - \eqref{eq:val} can be substantially simplified for some specific types of constraints. For instance, if we consider only the standard constraint \eqref{standard}, then 
\begin{equation*}
  \mathit{val}(\zeta)=\phi(\zeta+\gamma(\zeta)\1_n),
\end{equation*}
where $\gamma(\zeta)=n^{-1}(N-\1^T_n\zeta)$ for any $\zeta \in \Xi^{\mathrm{ex}}$, where $\1_n=(1,...,1)^T$.

Note that the heuristic evaluation of designs is chosen such that it depends only on the set $\Xi^{\mathrm{ap}}$ itself, not on the choice of the algebraic definition of $\Xi^{\mathrm{ap}}$. Furthermore, the excursions only depend on the ordering of approximate designs determined by the criterion $\phi$, not on the chosen ``version'' of the same criterion.

Clearly, there are many other methods of computing a local design evaluation in Algorithm \ref{Hugo}. For instance, it is possible to use some variant of the direct greedy method based on the relative change of $\phi$ with respect to a change in residual resources, similarly to \cite{KimYum}. These methods may allow for a more rapid construction of the excursion, nevertheless, they may also lose the above-mentioned invariance properties. Moreover, in our experiments with various modifications of move selection rules, we did not observe a significant increase in the quality of results.

\section{Examples} \label{sec:Examples}

In this section, we will apply Algorithm \ref{Hugo} to the most common situation in optimal design of experiments - computation of $D$-efficient experimental designs for regression models with independent homoscedastic errors, as described in the introduction. Although the chosen criterion is always the same, the selected optimization problems have very different sets of feasible designs. Our experience suggests that the feasible set has a more pronounced effect on the complexity of the optimization problem \eqref{GeneralProblem} than the choice of the criterion within the class of standard criteria used for optimal design. 

To demonstrate the universality of Algorithm \ref{Hugo}, we selected the same heuristic parameters in all examples (namely, $\mathit{back}_{\max}=16$ steps and $\mathit{n}_{\mathrm{round}}=9$ significant digits). We ran all computations for $\mathit{t}_{\max}=120$ seconds\footnote{For a specific optimization problem, we recommend experimenting with different values of $\mathit{t}_{\max}$ to estimate the time after which the heuristic does not lead to substantial improvements. Alternatively, the time-based stopping rule can be easily substituted by a stopping rule based on the number of iterations without improvement of the currently best design.}. To illustrate the statistical distribution of the quality of results and detect potentially difficult instances of the optimization problems, we used a set of $10$ independent initial designs generated by a random sequence of forward steps starting from $\xi^{(0)}$. In Example \ref{Dblock}, we used the \verb!R! computing environment\footnote{The reason is that a competing method for the problem of Example \ref{Dblock} is also written in R, i.e., we can provide a fair comparison.}, and in Examples \ref{sec:Rods}, \ref{sec:Work} we used an implementation of the general Algorithm \ref{Hugo} in \verb!Matlab!. The codes can be found at
\begin{center}
 \verb!www.iam.fmph.uniba.sk/design/!
\end{center}
All examples were computed on a 64 bit Windows 7 system running an Intel Core i5-2400 processor at 3.10 GHz with 4GB of RAM. 

\subsection{Designs for a block model with a constraint on the number of blocks and on the number uses of idividual treatments} \label{Dblock}

Consider a block model with $N$ blocks of size two and $v$ treatments. More precisely, assume that the independent observations $Y_1,...,Y_N$ satisfy
\begin{equation*}
 E(Y_j)=\tau(t_1(j))-\tau(t_2(j)), \: j \in \{1{:}N\},
\end{equation*}
where $t_1(j), t_2(j) \in \{1{:}v\}$ are the treatments selected for the $j$-th block, with effects $\tau(t_1(j)), \tau(t_2(j))$, and $\mathrm{Var}(Y_j)=\sigma^2 < \infty$, $j \in \{1{:}N\}$. An experimental design is given by a selection of treatments $t_1(j)$ and $t_2(j)$ to be compared in the $j$-th block, for all $j \in \{1{:}N\}$. Optimal designs for this model have been applied in two-channel microarray experiments (e.g., \cite{Wit}, \cite{Bailey}) and elsewhere.

In this setting, the design space can be viewed as the set of all possible pairs of treatments, i.e., $\mathfrak{X}=\{(1,2),(1,3),\dots,(v-1,v)\}$, which can be indexed by
\begin{equation*}
  \iota(t_1,t_2)=t_2-v+t_1v-(t_1^2+t_1)/2
\end{equation*}
for all $1 \leq t_1 < t_2 \leq v$. The problem of the so-called $D$-optimal block designs is then equivalent to the standard $D$-optimal design problem as described in the introduction, with $m=(v-1)$-dimensional regressors
\begin{equation*}
\f_{\iota(t_1,t_2)}=[\I_{v-1},\0_{v-1}](\e_{t_1}-\e_{t_2})
\end{equation*}
for all $(t_1,t_2) \in \mathfrak{X}$, cf. \cite{SagnolHarman}.

In this example, the aim is to demonstrate that Algorithm \ref{Hugo} performs well under the standard constraint \eqref{standard}, i.e., if the only restriction is not to exceed the given number $N$ of blocks. 

We implemented Algorithm \ref{Hugo} in the environment \verb!R! and used it to compute $D$-efficient designs for $v=16$ treatments and $N=15,\dots,120$ blocks. We then compared the designs with the results of a simulated annealing procedure $od$ implemented in \verb!R! package ``smida'' (see \cite{WNK}), with parameters $criterion="D"$, $dye=\mathrm{FALSE}$, and the number $n.iter$ of iterations chosen such that the computation time is approximately $t_{\max}=120$ seconds\footnote{The main application area of Algorithm \ref{Hugo} is computing efficient designs under non-standard constraints. Therefore, we did not perform a detailed comparison of Algorithm \ref{Hugo} with the vast number of other known methods applicable to computing optimal design under the standard constraint.}.

\begin{figure*}
\begin{centering}
\includegraphics[width=\textwidth]{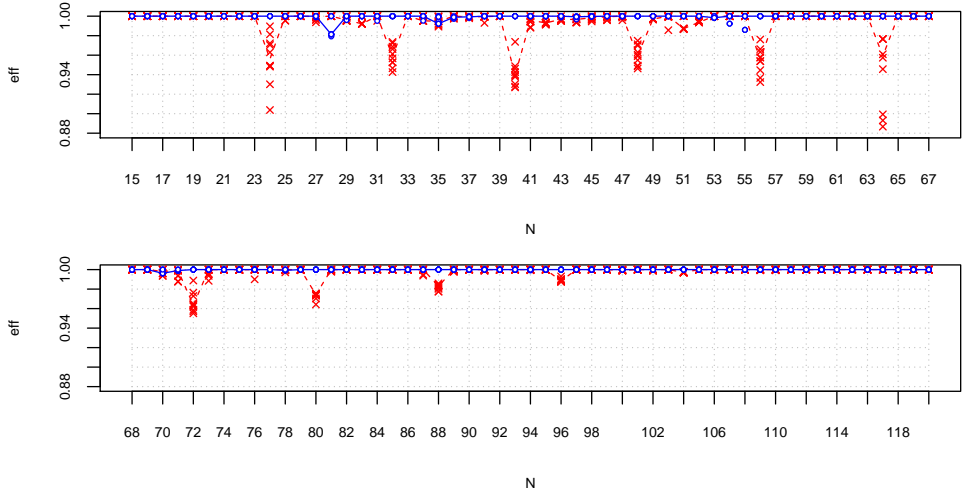}
\caption{Numerical results for the model from Subsection \ref{Dblock}. The horizontal axis corresponds to the number $N$ of blocks. The vertical axis corresponds to the $D$-efficiencies of the exact designs obtained by Algorithm \ref{Hugo} (circles) and the $D$-efficiencies of the exact designs obtained by a simulated annealing procedure form \cite{WNK} (crosses). All efficiencies are computed relative to the best exact design found by any of the methods. The solid line connects the medians of the sets of results of Algorithm \ref{Hugo}. The dashed line connects the medians of the sets of results produced by the competing simulated annealing method.
}
\label{releff}
\end{centering}
\end{figure*}

Figure \ref{releff} shows that Algorithm \ref{Hugo} systematically produced either the same or better results than the simulated annealing method (with small exceptions for $N=28, 36, 53, 54, 55, 78$). The numerical results suggest that the simulated annealing procedure has difficulties if $N$ is a multiple of $8$.

Any block design with blocks of size two can be represented by a ``concurrence'' graph with $v$ vertices and $N$ possibly multiple edges, such that the endpoints of edges correspond to the treatments used in the same blocks (e.g., \cite{Cameron}). Kirchhoff's matrix tree theorem implies that $\phi^m_D(\xi)$ is equal to the number of spanning trees of the concurrence graph of design $\xi$. Thus, the problem of $D$-optimal designs for this specific statistical model is equivalent to the problem of $t$-optimal graphs, that is, the concurrence graph of the $D$-optimal design maximizes the number of spanning trees in the class of graphs with fixed number of vertices and edges.

For some numbers $N$ and $v$, the $D$-optimal designs (or $t$-optimal graphs) are known theoretically (see \cite{Cheng}, \cite{Gaffke}, \cite{Petingi}). For instance, it is known that a complete almost-regular multipartite graph is $t$-optimal among all simple graphs with the same numbers of vertices and edges. The number of spanning trees for a complete multipartite graphs with $v$ vertices and $p \geq 2$ partitions of sizes $k_1,...,k_p$ is given by (\cite{Austin}, \cite{Lewis})
\begin{equation*}
\pi(v, k_1, k_2, \dots, k_p)=v^{p-2}\prod_{j=1}^{p}(v-k_j)^{k_j-1}.
\end{equation*}
With this formula, we can calculate the optimal value of the $D$-optimality criterion for $v=16$ and $N$=$64$, $85$, $96$, $102$, $106$, $109$, $112$, $113$, $\dots$, $120$. It turns out that Algorithm \ref{Hugo} consistently finds the theoretically $D$-optimal designs for all of these values of $N$.

\begin{figure*}
\begin{minipage}[b]{0.45\textwidth}
\centering
\includegraphics[width=\textwidth]{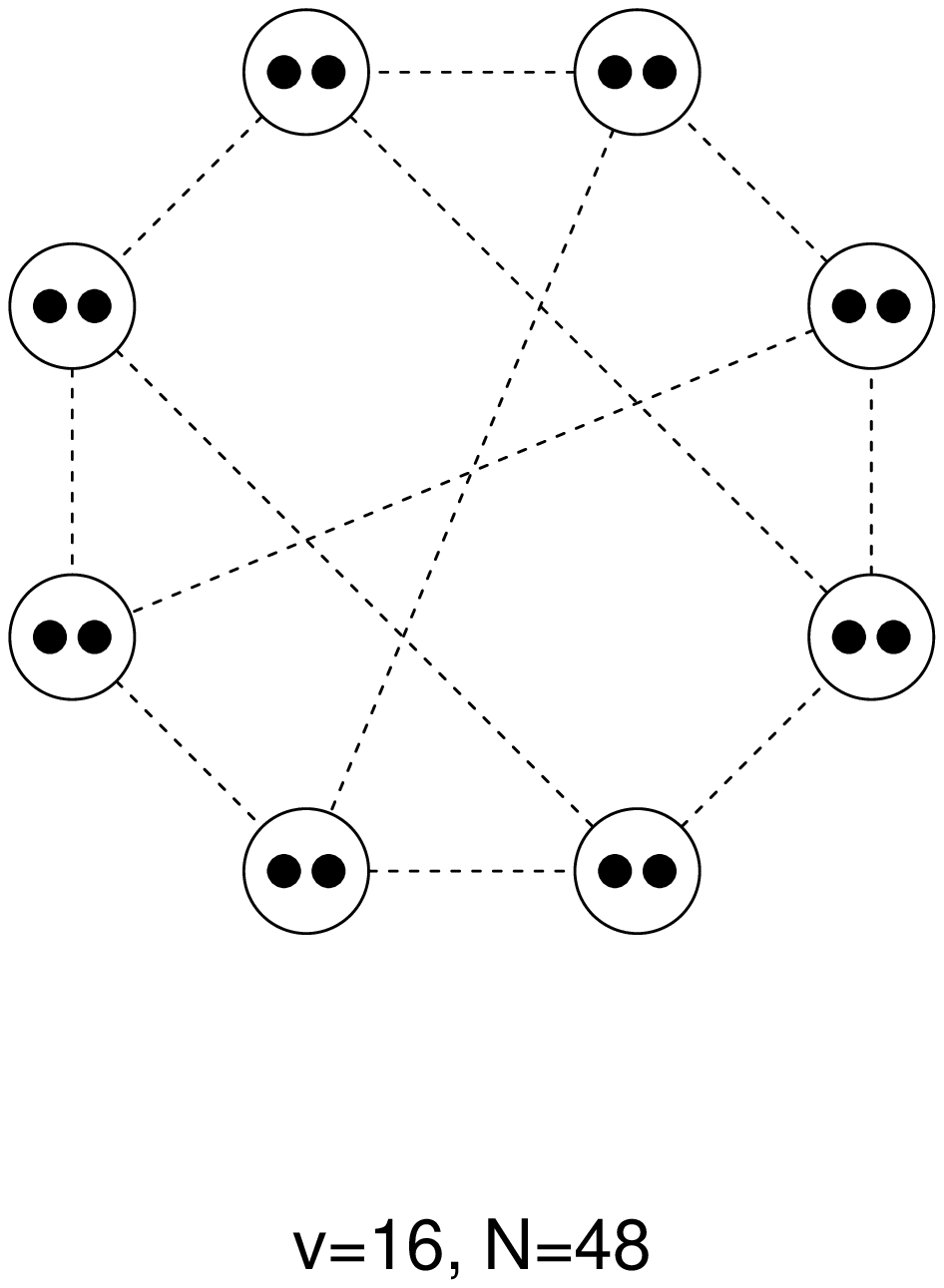}
\end{minipage}
\begin{minipage}[b]{0.45\textwidth}
\centering
\includegraphics[width=\textwidth]{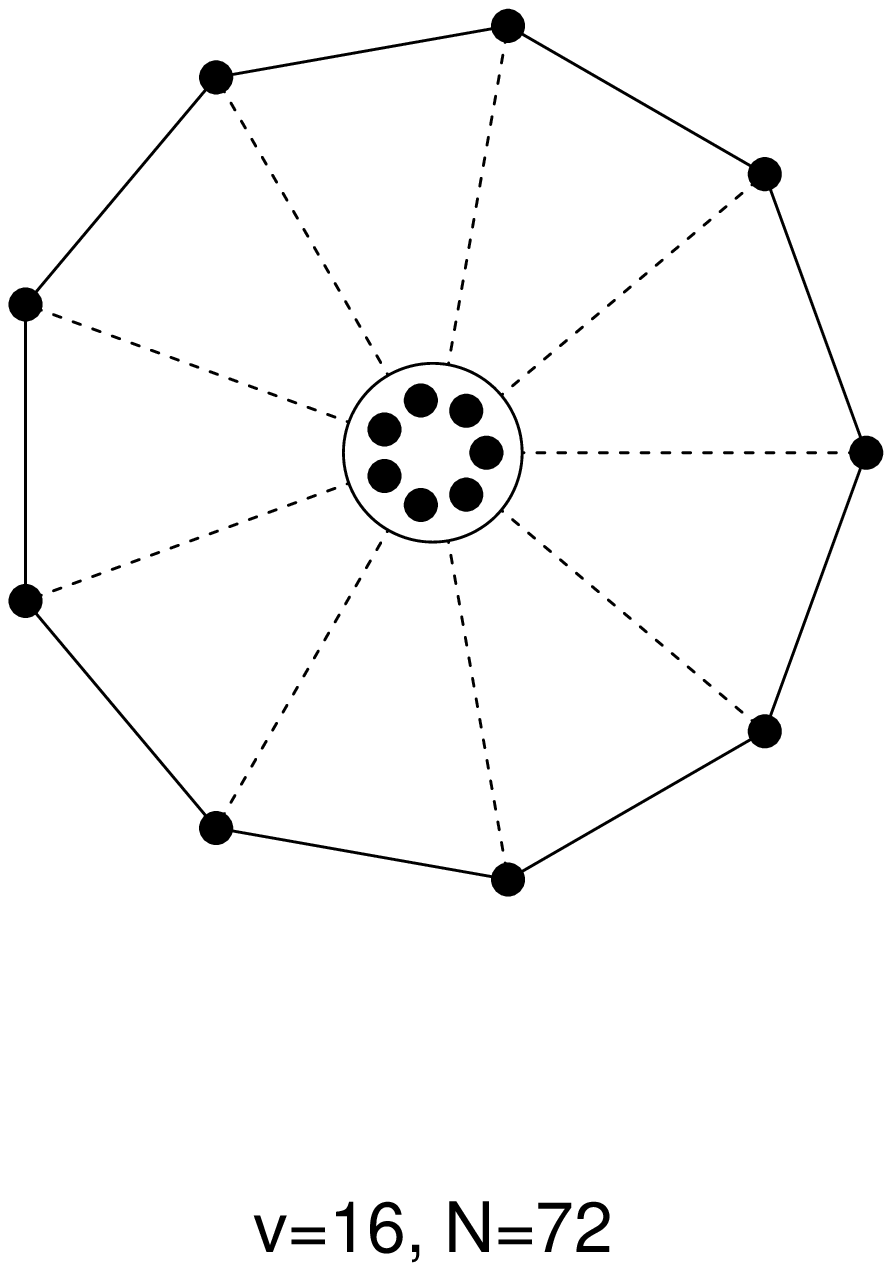}
\end{minipage}\\
\begin{minipage}[b]{0.45\textwidth}
\centering
\includegraphics[width=\textwidth]{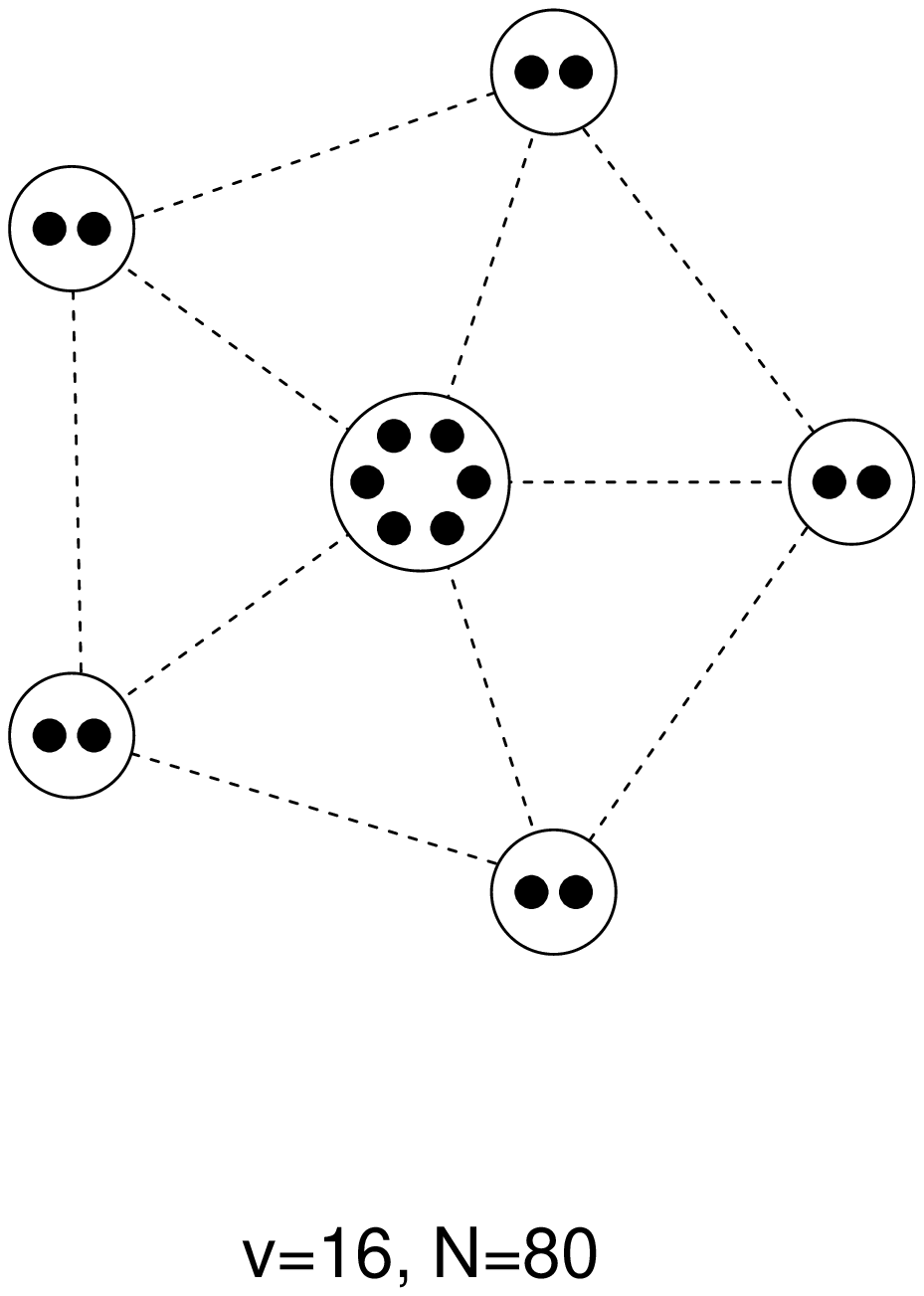}
\end{minipage}
\begin{minipage}[b]{0.45\textwidth}
\centering
\includegraphics[width=\textwidth]{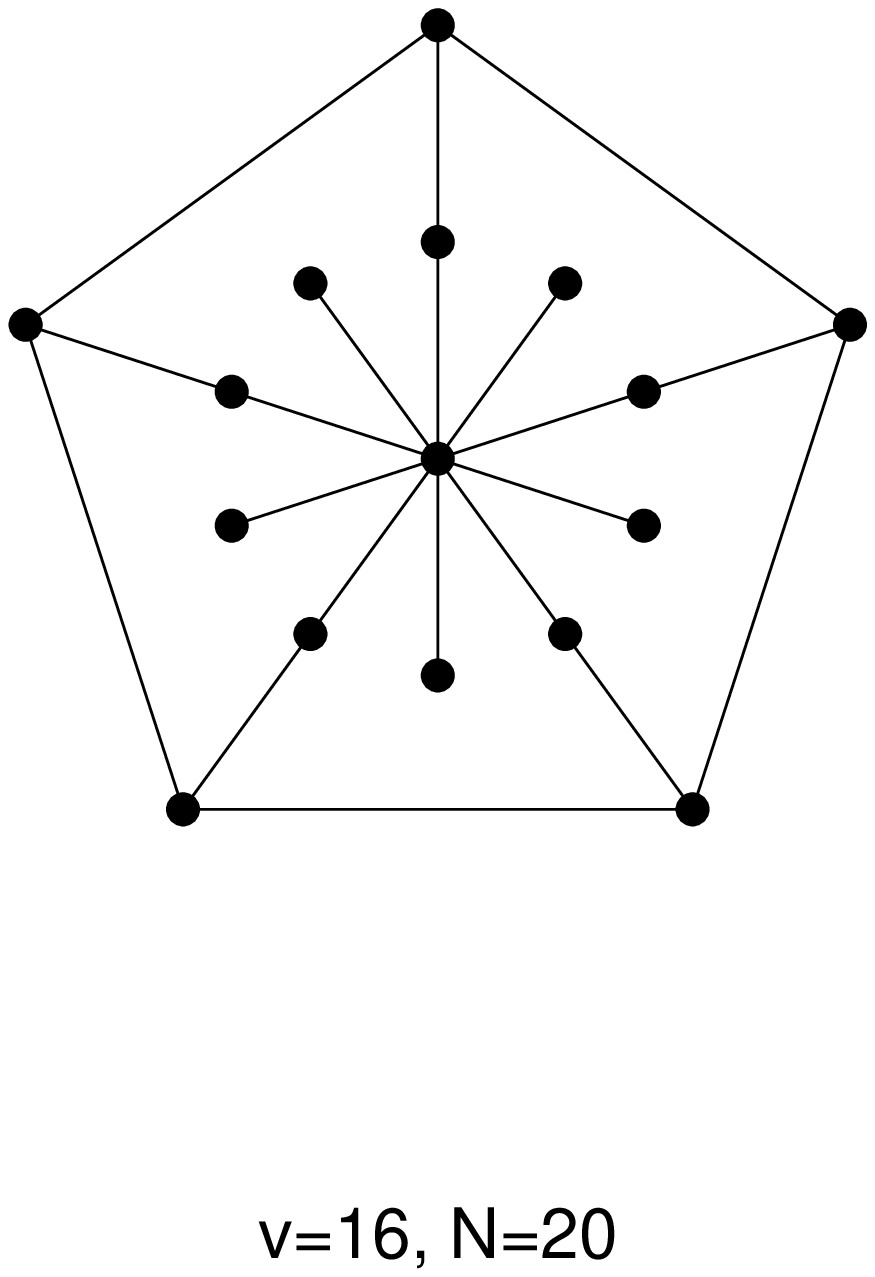}
\end{minipage}
\caption{The first three figures depict the concurrence graphs of the $D$-efficient designs obtained by Algorithm \ref{Hugo} for $v=16$ and $N=48,72,80$. The dashed lines represent complete multipartite subgraphs $K_{2,2}$ (for $N=48$), $K_{1,7}$ (for $N=72$) and $K_{2,2}$, $K_{2,6}$ (for $N=80$). The last graph illustrates treatment pairings for $20$ of the $65$ blocks of the $D$-efficient design with constraints on the numbers of uses of each treatment. See Subsection \ref{Dblock} for details.}\label{strong regular}
\end{figure*}

Compelling candidates for $D-$optimal designs are those that are represented by strongly regular graphs (srg), because of their high degree of symmetry. The imprimitive strongly regular graphs are either disconnected graphs or complete multipartite graphs with the partitions of the same size ($D-$optimal, as mentioned above). For $v=16$ vertices there exist four primitive strongly regular graphs  (see \cite{Colbourn},  Chapter VII.11). One of them, srg$(16,10,6,6)$, that is, the Clebsh graph with $N=40$ edges, was obtained by our heuristic, and we conjecture that it is $D$-optimal. However, the remaining three of the strongly regular graphs\footnote{Namely, the Shrikhande graph srg$(16,6,2,2)$ with $N=48$ edges, the complement of the Shrikhande graph srg$(16,9,4,6)$ with $N=72$ edges, and the complement of the Clebsch graph srg$(16,5,0,2)$ with $N=80$ edges.} are \emph{not} $D$-optimal; their efficiencies compared to the designs found by Algorithm \ref{Hugo} are $98.65\%$, $99.68\%$, and $99.61\%$, respectively. In Figure \ref{strong regular}, we depict the concurrence graph representation of the designs obtained by Algorithm \ref{Hugo} for $v=16$ and $N=48,72,80$. Interestingly, all these graphs  contain a large number of complete bipartite subgraphs.
\bigskip

To illustrate the possibilities of Algorithm \ref{Hugo} that go beyond the scope of the ``smida'' package, assume that we have no explicit limit on the number $N$ of blocks, but we do have upper limits on the replication numbers of individual treatments. Specifically, assume that $5$ treatments can be used at most $4$ times, $5$ other treatments at most $5$ times, another $5$ treatments at most $6$ times and one (say, standard) treatment at most $56$ times. These experimental restrictions can be formalized as resource constraints with $k=16$ inequalities, consumption coefficients $a_{r\iota(x_1,x_2)}=1$ for all $r \in \{1:16\}$ and all $x_1,x_2$ such that $r \in \{x_1,x_2\}$, and limits $b=(4,4,4,4,4,5,5,5,5,5,6,6,6,6,6,56)^T$. In $t_{\max}=120$ seconds, Algorithm \ref{Hugo} consistently produced a design with $N=65$ blocks that can be divided into two groups of blocks. The first group consists of $20$ blocks illustrated in the last graph of Figure \ref{strong regular}. The second group consists of $45$ blocks that compare each of the first $15$ treatments three times against the treatment $16$ (i.e., a star design replicated $3$ times).

\subsection{Designs for a quadratic model with simultaneous marginal and cost constraints} \label{sec:Rods}

Consider the $D$-optimal design problem for sintering uranium pellets that are to be used as a fuel in nuclear plants, as discussed in \cite{MartinMartin}. The explanatory variables represent the ``initial density'' ($x_1$) and the ``percentage of additive $U_3O_8$'' ($x_2$). The statistical model relating response  and the explanatory variables is assumed to be the full quadratic linear regression model with independent homoscedastic errors determined by the regressors 
\begin{equation}\label{modeluran}
 \f_{\iota(x_1,x_2)}=(1,x_1,x_2,x_1^2,x_2^2,x_1 x_2)^T.
\end{equation} 
 In \eqref{modeluran}, it is assumed that $(x_1,x_2)$ lies in $\mathfrak{X}=\{94.9,95.1,95,2,...,96.7\}\times \{0,10,20\}$, and the indices of the regression vectors are given by $\iota(94.9,0)=1$, $\iota(94.9,10)=2$, $\iota(94.9,20)=3$ and
\begin{equation*} 
 \iota(x_1,x_2)=30(x_1-94.9)+x_2/10-2
\end{equation*}
for all $(x_1,x_2) \in \mathfrak{X}$ such that $x_1 \geq 95.1$. Since the value 95.0 is missing from the factor levels of $x_1$, the design space $\mathfrak{X}$ has 54 points.

The nature of the experiment requires marginal constraints on the variable $x_1$ representing available experimental material (uranium rods). If we denote the required marginal sums by $b_1,\ldots,b_{18}$, the constraints on a feasible design $\xi$ are
\begin{equation*}
\xi_{3r-2}+\xi_{3r-1}+\xi_{3r} \leq b_r,\ r \in \{1{:}18\},
\end{equation*}
where ($b_1,\ldots,b_{18}$)=($1$, $3$, $14$, $59$, $52$, $29$, $25$, $32$, $36$, $29$, $36$, $38$, $12$, $10$, $8$, $2$, $3$, $3$).

Furthermore, we suppose that one percent of the additive costs one price unit (cf. \cite{HarmanFilova}, \cite{SagnolHarman}) and the financial resources of the experimenter are limited. Therefore, we solved the problem with additional constraints of the form
\begin{equation*}
10\sum_{r=1}^{18} \xi_{3r-1} + 20\sum_{r=1}^{18} \xi_{3r} \leq B,
\end{equation*}
where $B$ is a maximum possible cost of the experiment. We varied the maximum cost from $1100$ to $3900$ price units with a step $50$ and, for each $B$, we used Algorithm \ref{Hugo} to compute $10$ exact designs maximizing the criterion of $D$-optimality.

\begin{figure*}
\begin{centering}
	\includegraphics[width=\textwidth]{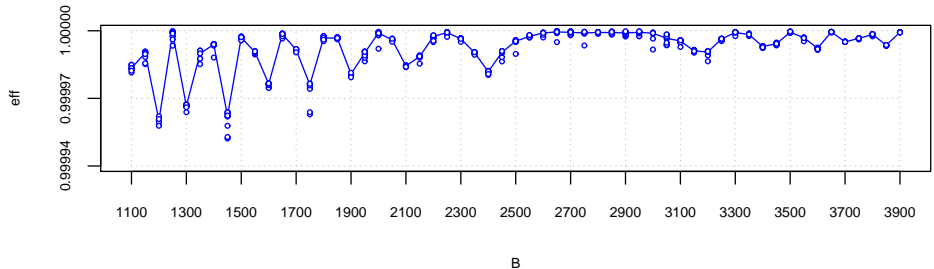}
	\caption{Numerical results for the model from Subsection \ref{sec:Rods}. The horizontal axis corresponds to the cost limit $B$. The vertical axis corresponds to the $D$-efficiencies of the exact designs obtained by Algorithm \ref{Hugo} relative to the approximate $D$-optimal designs. The line connects the medians of the sets of results of Algorithm \ref{Hugo}.}
	\label{tyce}
	\end{centering}
\end{figure*}

To express the quality of the resulting designs, we computed their $D$-efficiencies relative to the approximate $D$-optimal designs obtained by maxdet programming (see \cite{Vandenberghe}).  Figure \ref{tyce} shows that in all cases the efficiencies were higher than $99.99\%$. Note that the efficiencies relative to the approximate optimal designs represent lower bounds on the efficiencies relative to the (unknown) perfectly optimal exact designs. Additionally, the results are very stable in spite of the completely random selection of initial designs.

There are two mathematical programming methods that can be applied to the constrained problem from this subsection. The method based on integer quadratic  programming (IQP; \cite{HarmanFilova}) is often fast and simple to use, but, for this particular problem, it tends to produce worse results than Algorithm \ref{Hugo}. The approach based on mixed integer second order cone programming (MISOCP; \cite{SagnolHarman}) gives more efficient designs than IQP, but its results are still slightly worse than the results of Algorithm \ref{Hugo}, even if the MISOCP solver is run for a very long time.

The IQP and the MISOCP methods are more complex than Algorithm \ref{Hugo}, often provide worse designs, and require an advanced integer programming solver. Nevertheless, note that they can be applied under more general linear constraints than Algorithm \ref{Hugo}. Moreover, the MISOCP method provides a non-trivial lower bound on the efficiency of the resulting design. 

\begin{figure*}
\begin{centering}
	\includegraphics[width=\textwidth]{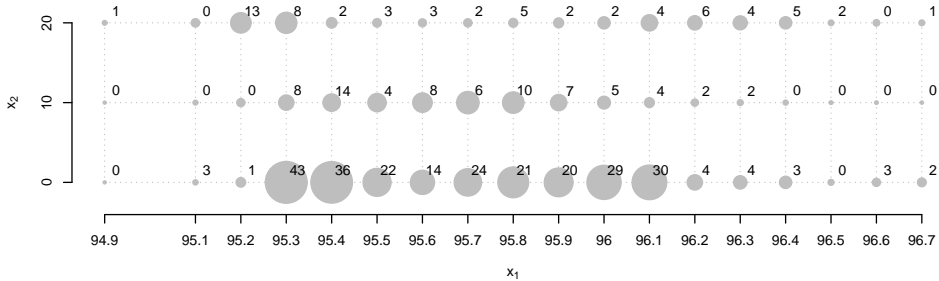}
	\caption{The marginally constrained $D$-optimal approximate design and the best marginally constrained exact design found by Algorithm \ref{Hugo}, with the additional cost constraint $B=1965$. The relative weights of the $D$-optimal approximate design are denoted by the sizes of the grey discs. The integer numbers represent the exact design found by Algorithm \ref{Hugo}. Note that the exact design is very close to the optimum since its $D$-efficiency is higher than $99.92\%$. See Subsection \ref{sec:Rods} for details.}
	\label{tycecost}
\end{centering}
\end{figure*}

To show a concrete example of a marginally and cost constrained design, we chose $B=1965$; see Figure \ref{tycecost} for the result. The exact design obtained in \cite{SagnolHarman} by MISOCP has $D$-efficiency of about $1-10^{-5}$ relative to the exact design obtained by Algorithm \ref{Hugo}.

\subsection{Designs for a non-linear regression model with simultaneous direct and cost constraints} \label{sec:Work}

The third example is taken from \cite{Wright}. Suppose that we wish to find the best sequence of sampling times for a model relating time and internal concentrations of fluoranthene in an organism. The mean internal concentration at time $t$ of the experiment is given by
\begin{equation}\label{wwmodel}
\mu_t(\theta_1,\theta_2)=\frac{\theta_1}{\theta_2}\left(e^{-\theta_2\max\{t-72,0\}}-e^{-\theta_2 t} \right),
\end{equation}
where $\theta_1$ and $\theta_2$ are parameters corresponding to the constant uptake and elimination rates. The experiment will be initiated at a starting time $s$ and all observations need to be performed within the following $144$ hours. Hence, an appropriate designs space is $\mathfrak{X}=\{0,1,...,144\}$ and the elements $t \in \mathfrak{X}$ represent the time (in hours) elapsed from $s$.

The model in consideration is non-linear, therefore we will compute the locally $D$-optimal designs (see, e.g., \cite{PronzatoPazman}). To this end, we need to linearise the model in some fixed parameters $\vartheta_1$ and $\vartheta_2$. Since the model is linear in $\theta_1$, the choice of $\vartheta_1$ is irrelevant. For $\theta_2$, we will select the nominal value $\vartheta_2=0.2381$ suggested in \cite{Wright} by earlier experiments.

Thus, we will consider the $D$-optimal design problem for the linear regression model with independent homoscedastic errors and two-dimensional regressors 
\begin{equation*}
 \f_{\iota(t)}=\nabla \mu_t(\vartheta_1,\vartheta_2),
\end{equation*} 
where $t \in \mathfrak{X}$ is the design point corresponding to the time of the observation, $\iota(t)=t+1$ is the index of the design point, and $\nabla$ denotes the gradient. 

The experiment requires observations at $t=0$, $t=72$, and $t=144$ hours of the experiment. Thus, the design $\xi^{(0)}$ to be augmented satisfies 
$\xi^{(0)}_1=\xi^{(0)}_{73}=\xi^{(0)}_{145}=1$, and  $\xi^{(0)}_i=0$ for all $i\in \{1{:}145\}\setminus\{1,73,145\}$.

In accord with \cite{Wright}, we also assume that the experimental budget of the practitioner is limited by $B=13$ price units. Moreover, the sampling costs vary throughout the week. For a starting time $s$, we can divide the design space $\mathfrak{X}$ as
\begin{equation*}
\mathfrak{X}=\mathfrak{X}^{s}_1 \cup \mathfrak{X}^{s}_2 \cup \mathfrak{X}^{s}_{1.5},
\end{equation*}
where $\mathfrak{X}^{s}_{1}$ denotes the sampling times with regular hourly wage on weekdays (8am - 5pm, Monday to Friday), $\mathfrak{X}^{s}_{2}$ denotes the sampling times with double wage on the weekend (7pm Friday - 6am Monday), and $\mathfrak{X}^{s}_{1.5}$ denotes the sampling times with $1.5$ of regular wage (all other times). Hence, if $c$ is the cost of taking a sample at a time with a regular hourly wage, the cost of taking one sample in $\mathfrak{X}^{s}_{2}$ will be $2c$ and the cost of taking a sample in $\mathfrak{X}^{s}_{1.5}$ will be $1.5c$. We are interested in finding optimal designs that do not exceed the budget $13c$. Additionally, we can perform at most one observation in each design point.

Formally, the constraints required by the experimental set-up can be expressed as:
\begin{equation*}
 \sum\limits_{i\in \mathfrak{X}^{s}_{1}}\xi_i+1.5\sum\limits_{i\in \mathfrak{X}^{s}_{1.5}}\xi_i+2\sum\limits_{i\in \mathfrak{X}^{s}_{2}}\xi_i \leq 13, 
\end{equation*}
and $\xi_i\in\{0,1\}$, $i \in \{1{:}145\}$, for any feasible design $\xi$.

\begin{figure*}
\begin{centering}
	\includegraphics[width=\textwidth]{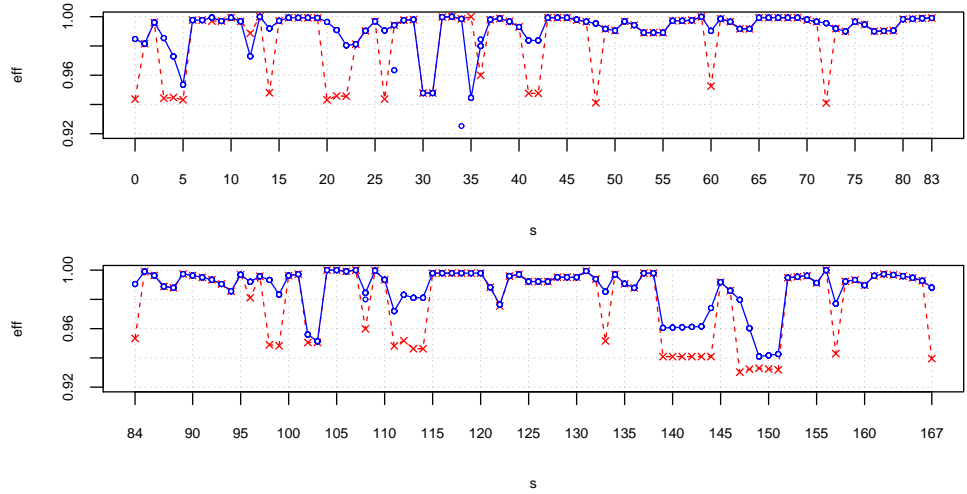}
	\caption{Numerical results for the model from Subsection \ref{sec:Work}. The horizontal axis corresponds to the starting time $s$ of the experiment. The vertical axis corresponds to the $D$-efficiencies of the exact designs obtained by Algorithm \ref{Hugo} (circles) and the $D$-efficiencies of the exact designs obtained by the heuristic from \cite{Wright} (crosses). All efficiencies are computed relative to the $D$-optimal approximate designs. The solid line connects the medians of the sets of results of Algorithm \ref{Hugo}. The dashed line connects the results of the heuristic from \cite{Wright}.
}
	\label{workweek}
\end{centering}
\end{figure*}

We have used Algorithm \ref{Hugo} as well as the heuristic from \cite{Wright}\footnote{We used the \texttt{Matlab} code provided on the web page of the authors of \cite{Wright}.} to compute $D$-efficient designs for starting times $s=0,\ldots,167$. To express the quality of the obtained designs, we have evaluated their efficiencies relative to the locally $D$-optimal approximate designs computed by maxdet programming; see Figure \ref{workweek}. Similarly to the previous examples, the results of Algorithm \ref{Hugo} are very stable; all random restarts resulted in the same design except for $s=27, 34, 36, 108$. Moreover, all $1680$ results of Algorithm \ref{Hugo} were the same or better\footnote{Note that the procedure from \cite{Wright} produced the results significantly faster than in $t_{\max}=120s$, however, because of its deterministic nature, it cannot further improve its results.} the corresponding results from \cite{Wright} with the following exceptions: $10$ results for $s=12$, $3$ results for $s=27$, $1$ result for $s=34$, and $10$ results for $s=35$. In the most problematic case of $s=35$, the $D$-efficiency of all $10$ results of Algorithm \ref{Hugo} is only $94.46\%$ relative to the design found by the heuristic from \cite{Wright}. Nevertheless, our computational experiments show that for $s=35$ Algorithm \ref{Hugo} detects the optimal design after $200$ to $300$ seconds, depending on the initial design.

\begin{figure*}
\begin{centering}
	\includegraphics[width=\textwidth]{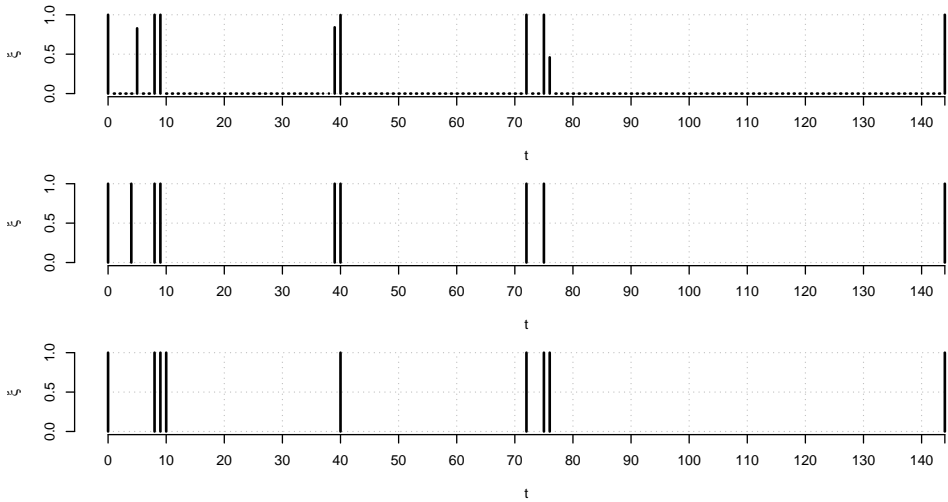}
	\caption{The $D$-optimal approximate design (upper panel) and two exact designs (middle and lower panels) for the model from Subsection \ref{sec:Work} with starting time $s=72$. The graph on the middle panel represents the exact design obtained in \cite{Wright} and the graph on the lower panel represents the exact design obtained by Algorithm \ref{Hugo}.}
 \label{ww73}
\end{centering}
\end{figure*}

Concrete examples of the experimental designs (for the starting time $s=72$) are depicted in Figure \ref{ww73}. For this case the relative efficiency of the design found by Algorithm \ref{Hugo} with respect to the approximate $D$-optimal design is $99.56\%$, whereas for the design obtained by \cite{Wright}, the $D$-efficiency is $94.09\%$.

\section{Conclusions}

We showed that the resource constraints \eqref{resconst} cover many types of experimental design restrictions, and that the optimal design problems associated with these restrictions can be efficiently solved by a common heuristic. For simplicity, we computed the numerical examples for the criterion of $D$-optimality, but it is straightforward to apply the heuristic to any monotonic criterion. Moreover, the algorithm can be as easily applied to statistical models different from the standard regression.

There are many variants of the proposed heuristic that could enhance its performance in specific situations. Besides alternative choices of initial designs, formulas for designs' characteristic attributes and local heuristic evaluations, it might also be possible to improve the efficiency of the heuristic by a different notion of a ``failed'' excursion, or  variations in the manipulation with the list $\V$. Since the heuristic is based on forward and backward steps, the speed of the execution could also be enhanced using the update formulas from \cite{Atkinson}, Chapter 12.  

Clearly, multitudes of nature-inspired optimization heuristics, such as physical, evolutionary and swarm algorithms are also applicable to solving problems of type \eqref{GeneralProblem}, either directly or using a penalty approach to take the constraints into account (cf., e.g., \cite{Kirkpatrick}, \cite{MichalewiczFogel}, \cite{Eberhart}, cf. also \cite{Haines}, \cite{Montepiedra}, \cite{Mandal} for the applications of these methods to the standard optimum design problem). However, these methods usually require a large amount of programmer's experience, numerical experimentation and fine tuning of parameters to fit the specific properties of the optimization problem at hand.

Thus, besides introducing the general resource constrained problem in the area of experimental design, a secondary aim of this paper was to provide a simple, universal, yet reasonably efficient benchmark method for testing more advanced techniques that might be developed in the future.

\section*{Acknowledgement}
The research was supported by the VEGA 1/0163/13 grant of the Slovak Scientific Grant Agency.


\begin{thebibliography}{50}

\bibitem{Atkinson} Atkinson, A.C., Donev, A.N., Tobias, R.D.: Optimum Experimental Designs, With SAS. Oxford University Press (2007)

\bibitem{Austin} Austin, T.L.: The enumeration of point labelled chromatic graphs and trees. Canad. J. Math. 12, 535-545 (1960)

\bibitem{EYSM} Bachrat\'{a}, A., Harman, R.: A stochastic optimization method for constructing optimal block designs with linear constraints. Proceedings from the European Young Statisticians Meeting, Osijek, 2013 (2014)

\bibitem{Bailey-book} Bailey, R.A.: Design of Comparative Experiments. Cambridge Series in Statistical and Probabilistic Mathematics. Cambridge University Press (2008)

\bibitem{Bailey} Bailey, R.A.: Designs for two-colour microarray experiments. J. Roy. Stat. Soc. C Appl. Stat. 56(4), 365-394 (2007)

\bibitem{Cameron} Bailey, R.A., Cameron, P.J.: Combinatorics of optimal designs. Surveys in Combinatorics - London Mathematical Society Lecture Note Series 365, 19-73 (2009)

\bibitem{Cheng} Cheng, C.S.: Maximizing the total number of spanning trees in a graph: two related problems in graph theory and optimum design theory. J. Combin. Theor. B 31, 240-248 (1981)

\bibitem{Chern} Chern, M.S.: On the computational complexity of reliability redundancy allocation in a series system. Oper. Res. Lett. 11, 309-315 (1992)

\bibitem{CookThibodeau} Cook, R.D., Thibodeau, L.A.: Marginally restricted D-optimal designs. J. Am. Stat. Assoc. 75(370), 366-371 (1980)

\bibitem{CookFedorov} Cook, R.D., Fedorov, V.V.: Constrained Optimization of Experimental Design. Statistics 26, 129-178 (1995)

\bibitem{Colbourn} Colbourn, C.J., Dinitz, J.H.: CRC Handbook of Combinatorial
Designs (2nd ed.). CRC Press (2007)

\bibitem{Dykstra} Dykstra, O.: The augmentation of experimental data to maximize $|\mathrm{X'X}|$. Technometrics 13, 682-688 (1971)

\bibitem{Eberhart} Eberhart, R.C., Shi, Y., Kennedy, J. : Swarm Intelligence. Morgan Kaufmann (2001)

\bibitem{Elfving} Elfving, G.: Optimum allocation in linear regression theory. Ann. Math. Stat. 23(2), 255-262 (1952)

\bibitem{Fedorov82} Fedorov, V.V.: Optimal design with bounded density: Optimization algorithms of the exchange type. J. Stat. Plann. Infer. 22, 1-13 (1982)

\bibitem{FedorovHackl} Fedorov, V.V., Hackl, P.: Model-Oriented Design of Experiments. Springer (1997)

\bibitem{Gaffke} Gaffke, N.: D-optimal block designs with at most six varieties, J. Stat. Plann. Infer. 6, 183-200 (1982)

\bibitem{GloverLaguna} Glover, F., Laguna, M.: Tabu search. Springer (1999)

\bibitem{Haines} Haines, L.M.: The application of the annealing algorithm to the construction of exact optimal designs for linear-regression models. Technometrics 29, 439-447 (1987)

\bibitem{Jones} Goos, P., Jones, B.: Optimal Design of Experiments: A Case Study Approach. John Wiley \& Sons, New York (2011)

\bibitem{JungYum} Jung, J.S., Yum, B.J.: Construction of exact D-optimal designs by tabu search. Computational Statistics \& Data Analysis 21, 181-191 (1996)

\bibitem{Kellerer} Kellerer, H., Pferschy, U., Pisinger, D. : Knapsack Problems. Springer (1993)

\bibitem{Harman14} Harman, R.: Multiplicative Methods for Computing D-Optimal Stratified Designs of Experiments. J. Stat. Plann. Infer. 146, 82 - 94 (2014)

\bibitem{HarmanFilova} Harman, R., Filov\'{a}, L.: Computing efficient exact designs of experiments using integer quadratic programming. Comput. Stat. Data Anal. 71, 1159-1167 (2014)

\bibitem{Hochbaum} Hochbaum, D.S.: A nonlinear Knapsack problem. Oper. Res. Lett. 17(3), 103 - 110 (1995)

\bibitem{KimYum} Kim, J., Yum, B.: A heuristic method for solving redundancy optimization problems in complex systems. IEEE Trans. Reliab. 42(4), 572-578 (1993)

\bibitem{Kuo} Kuo, W., Prasad, V.R., Tillman, F., Hwang, C.: Optimal reliability design: fundamentals and applications. Cambridge university press (2001)

\bibitem{Lewis} Lewis, R.P.: The number of spanning trees of a complete multipartite graph. Discrete Math. 197/198(0), 537 - 541 (1999)

\bibitem{Mandal} Mandal, A., Wong, W.K., Yu, Y.: Algorithmic Searches for Optimal Designs, in: Handbook of Design and Analysis of Experiments, Chapman \& Hall/CRC (2014)

\bibitem{MartinMartin} Mart\'{i}n-Mart\'{i}n, R., Torsney, B., L\'{o}pez-Fidalgo, J.: Construction of marginally and conditionally restricted designs using multiplicative algorithms. Comput. Stat. Data Anal. 51(12), 5547-5561 (2007)

\bibitem{MichalewiczFogel} Michalewicz, Z., Fogel, D.B.: How to Solve It: Modern Heuristics. Springer (2008) 

\bibitem{Mitchell} Mitchell, T.J.: An Algorithm for the Construction of "D-Optimal" Experimental Designs. Technometrics 16(2), 203-210 (1974)

\bibitem{Montepiedra} Montepiedra, G., Myers, D., Yeh, A.B.: Application of genetic algorithms to the construction of exact D-optimal designs. Journal of Applied Statistics 25, 817-826 (1998)

\bibitem{ParkMontgomery} Park, Y., Montgomery, D.C., Fowler, J.W., Borror, C.M.: Cost-constrained G-efficient Response Surface Designs for Cuboidal Regions. Qual. Reliab. Eng. Int. 22(2), 121-139 (2006)

\bibitem{Pazman} P\'{a}zman, A.: Foundations of optimum experimental design. Reidel (1986)

\bibitem{Petingi} Petingi, L., Rodriguez, J.: A new technique for the characterization of graphs with a maximum number of spanning trees. Discrete Math. 244, 351-373 (2002)

\bibitem{PronzatoPazman} Pronzato, L., P\'{a}zman, A. : Design of Experiments in Nonlinear Models: Asymptotic Normality, Optimality Criteria and Small-sample Properties. Springer (2013)

\bibitem{Pukelsheim} Pukelsheim, F.: Optimal Design of Experiments. SIAM (2006)

\bibitem{R} R Development Core Team: R: A Language and Environment for Statistical Computing, Foundation for Statistical Computing. Vienna, Austria (2011)

\bibitem{SagnolHarman} Sagnol, G., Harman, R.: Computing exact D-optimal designs by mixed integer second order cone programming. arXiv preprint, arXiv:1307.4953 (2013)

\bibitem{Kirkpatrick} Schneider, J., Kirkpatrick, S.: Stochastic Optimization. Springer (2006)

\bibitem{TackVandebroek} Tack, V., Vandebroek, M.: Budget constrained run orders in optimum design. J. Stat. Plann. Infer. 124, 231-249 (2004)

\bibitem{Vandenberghe} Vandenberghe, L., Boyd, S., Wu, S.P.: Determinant maximization with linear matrix inequality constraints. SIAM J. Matrix Anal. Appl. 19, 499-533 (1998)

\bibitem{Welch82} Welch, W.J.: Branch-and-bound search for experimental designs based on D-optimality and other criteria. Technometrics 24(1), 41-48 (1982)

\bibitem{Welch84} Welch, W.J.:Computer-Aided Design of Experiments for Response Estimation. Technometrics 26(3), 217-224 (1984)


\bibitem{Wit} Wit, E., McClure, J.: Statistics for Microarrays. Wiley (2004)

\bibitem{WNK} Wit, E.,  Nobile, A., Khanin, R.: Near-Optimal Designs for Dual Channel Microarray Studies.  J. Roy. Stat. Soc. C Appl. Stat. 54(5), 817-830 (2005)

\bibitem{Wright} Wright, S.E., Sigal, B.M., Bailer, A.J.: Workweek Optimization of Experimental Designs: Exact Designs for Variable Sampling Costs. J. Agr. Biol. Environ. Stat. 15(4), 491-509 (2010)

\bibitem{Wynn} Wynn, H.P.: The sequential generation of D-optimum experimental designs. Ann.Math.Statist. 41, 1655-1664 (1970)


\end{thebibliography}
\end{document}